\documentstyle[emlines2]{article}
\begin{document}
\begin{center}
The Maximum of Potential Energy and Symmetries for Massive Galaxies and Stars.

{\ }

 Rashid Faizullin

{\ }
 Omsk State University, Omsk, Russia, faizulin$@$univer.omsk.su

{\ }

\end{center}

                    Abstract:
    The consideration of the N-body gravitational problem equations can
  give to us some class of boundary-value problems defined on the "beem's"
  construction. One can  considere it as weak or so-called finite element
  method's approximation with the linear  function's basis.  So there are
  time and coordinate dependent beems  as  distances between  points  and
  unknown coordinate functions which are defined on the beems. The possible
  sample from this class is simple Lalplas equation where potential energy
  are maximal. Some computer simulations can give to us symmetries on
  spherical coordinates (similar for well known mathematical structures -
  configurations like Pascal or Desargus configurations).

  But in other hand, the  consideration of the natural mass distribution
  can give to us evidence of the physical reality of the similar structures
  on the star sphere. Moreover, there are coupled invariancy for geometry
  and arithmetics.

{\ }

51P05, 85-XX

{\ }


{\ }

{\bf Introduction. \rm}

{\ }

 So, it is well known that there is a structure of young $O,B$ and $A0$
 stars. Herschel (1847) was the first who noted  and statistically
 convincing arguments were given by Guld (1879), that there is so called
 Guld zone. There was discussion of physical validity and stability-
 problem of so called $O,B$ associations and superassociations $[ 1 ]$,
 $[ 2 ]$, $[ 3 ]$. At the last time appeared the  results on the periodicity
 for the clusters of the galaxies  $[ 4 ]$, $[ 5 ]$.
 Possibly there is mathematical unity for some astrophenomena
 or in other words automodelity for mass distribution so
 J.Einasto wrote:
  ' If this reflects the distribution of all matter (luminous
   and dark), then there must exists some hitherto unknown process that
   produces regular structure on large scales' $[ 5 ]$.
 But word 'unknown' does not mean existence of really new physical law
 which take place only for large scales. We  hope that there are
 mathematical methods yet and order for distances smaller than
 cosmological sizes.
 We start our investigation from famous and old N-body problem $[ 6 ]$,
 then for assemblies of bodies with low velocities we can find coherent
 structures.

 But there is similar visual coherentness, moreover automodel symmetry
 for some scales on natural  coordinate systems ( spherical systems of
 coordinates related with momentum).

{\ }

{ \bf N-body equations for $\ddot x, \ddot y, \ddot z=0 $. \rm }

{\ }

  Let us consider N-body equations for N points with masses $M_j$
 on OXYZ coordinate system $[ 7 ]$:

 $$\ddot{u}_i =  (\sum_{j=1}^{j=N'}
 M_j(u_j -u_i)/ r_{ij}^3), u_i=x_i, y_i, z_i  \eqno(1)$$

  where $r_{ij}$ is distance between points i and j, sign ' noted
 so there is no case i=j.

 After decomposition  we can write:

 $$\ddot{u}_i = - u_i (\sum_{j=1}^{j=N'} M_j/r_{ij}^3)+ (\sum_{j=1}^{j=N'}
 u_j M_j/ r_{ij}^3), u_i=x_i, y_i, z_i \eqno(2)$$

 Note, that right parts of (2)  are  equal to product of
  the positive matrix $A(x,y,z)$  (by weak diagonal predominance) on the
 vector $w=(x_1,x_2,.. x_N$, $y_1,..y_N,....z_N)$.

  Then let us consider (2) as finite approximation of some
 boundary-value problem on the "beem's" mesh where lenghtess
 of beems or in other words distances between points must be
 solution depended.

 We'll try find form from our finite-difference equations or 'revers motion'
on well known transformations:

$$ \rho_t +(\rho \phi_x)_x +(\rho \phi_y)_y=0,  \phi_n =0 \eqno(3)$$
$$ v(x,y)(\rho_t +(\rho \phi_x)_x +(\rho \phi_y)_y)=0, \eqno(4)$$
$$\int_{\Omega} (v \rho_t -\rho \phi_x v_x -\rho \phi_y v_y)d\omega=0 \eqno(5)$$

  For right part we can find function $f(l)$  from:
 $$ Right Part=\int_L f(l) u_l v_l dl \eqno(6)$$
 where L is beem in the distinguished moment of the time
 and integration are going along directions between
 points,  $v(l)$ is test function.

 Then we can take finite basis of functions $v_i(l)$ (Figure 1)
 which related with points and which have finite region
 where $v(l)$ is not equal to zero (so-called finite elements $[8]$).
 In (3) we have:
$$ u(l)=\sum_{k=1}^{k=N} \alpha_k v_k(l) \eqno(7)$$
 After that we have that $f(l)=M_j/R^2$-- constant on every beem
 where $R=r_{ij}$.

 For left part we have:
 $$Left Part=2/K (\int_L v(l) u_{tt}/R - R/6 u_{ttl}v_ldl) \eqno(8)$$
 where $K$ is the number of the nodes which significant connected
 with the given one.

 In differential form and for the equal masses M we have:

 $$ Mu_{tt}/R +(MR/6 u_{ttl})_l= (M^2/R^2 u_l )_l \eqno(9)$$
 where we take $K=2$ without less of generality with respect to physical
 case.

  We can consider the row of  equations
  where one of them is (6) and others are:

 $$ M/R u_{tt} = (M^2/R^2 u_l )_l \eqno(10)$$
 $$ (M/Ru_l)_l=0 \eqno(11)$$

 where $R$ is equal to small number when we take by unit size of our
 constellation and gravitational interaction between points defined
 on finite distance which are less than  $MR$ for some natural $M$.

 Let us consider (11). We can see so there are spherical symmetry
 for form (11) and hence there are some hope for symmetries
 on spherical representations on 'natural' ( that it to say sperical)
 coordinates.

There are easy way for computer simulation.
Let us consider volume $G$ and points with equal masses where some points
(set $A$) are into the
volume $G$ and other points (set $B$) are outside the volume $G$. The
every point
is surrounded by an $P$ envelope of some (set $P$ is not equal $A$+$B$)
points. Note, near $G$ boundary $P_z$ have points from $A$ and $B$ too.

Let's define iterative motion for every point $z$ from volume $G$
into mass center of $P_z$. Then after some steps we will have solution
for (11).

We can see some  examples of simulation on Figures 2, 3, 4 where $P$
defined by choosen distance.

By the way --- it is simple and fast algorithm for Steiner problem $[ 10 ]$
 for two or three dimensional cases--- Figure 5.

{\ }

{ \bf Application \rm}

{\ }

Well known so equation  (11) seems as fully unacceptable for real masses and
formal solutions of (11) seems as very inconstant by time.

{\ }

But let us consider nearest and well known light galaxies --- M31, M81, M82, M83,
 M101, NGC 5128, Dwingeloo1 (data from nedwww.ipac.caltech.edu).

 We can see pictures --- Figures 6,7,8,9 which are similar for our
computer simulation.

{\ }

At next let us consider "histogram" for  natural stars as distribution
  on the star's sphere or in other words as the distribution
  on the one of the  planes:($l,b$), ($\alpha, \delta$) or
  ($\lambda, \beta$) on the one or two coupled rectangles of size
  $2\pi *\pi$ ( there are periodicity).
  Also there is physical value for star which can be correlated
  with possible order. It is so-called visual brightness or in the order
  words order which was
  related with some degree of the $M^2/R^{0.6}$ where $R$ is distance between
  star and view point $[ 11 ]$. Below are given the enumeration for 50
  lightest stars $[ 11 ]$:

{\ }

        1 $\alpha Cma $
        2 $\alpha Car $
        3 $\alpha Bool$
        4 $\alpha Lyr$
        5 $\alpha Cen $
        6 $\alpha Aur$
        7 $\beta Ori $
        8 $\alpha Cmi$
        9 $\alpha Ori$
        10 $\alpha Eri$
        11$\beta Cen$
        12 $\alpha Aql$
        13 $\alpha Cru$
        14 $\alpha Tau$
        15$\alpha Sco$
        16 $\alpha Vir$
        17 $\beta Gem$
        18 $\alpha Psa$
        19  $\beta Cru$
        20 $\alpha Cyg$
        21  $\alpha Leo$
        22 $\epsilon Cma$
        23 $\alpha Gem$
        24  $\lambda Sco $
        25 $\gamma Ori$
        26 $\gamma Cru$
        27  $\beta Tau$
        28 $\beta Car$
        29 $\epsilon Ori$
        30 $\alpha Gru$
        31 $\epsilon Uma$
        32 $\zeta Ori$
        33 $\alpha Uma$
        34 $\alpha Per$
        35 $\gamma Vel$
        36 $\epsilon Sgr$
        37 $\delta CMa$
        38 $\eta Uma$
        39 $\epsilon Car$
        40 $\theta Sco$
        41 $\beta Aur$
        42 $\gamma Gem$
        43 $\alpha Tra$
        44 $\delta Vel$
        45 $\alpha Pav$
        46 $\beta CMa$
        47 $\alpha Hya$
        48 $o Cet$
        49 $\alpha Ari$
        50 $\alpha UMi$

{\ }

 As an example, let us consider the stars with numbers
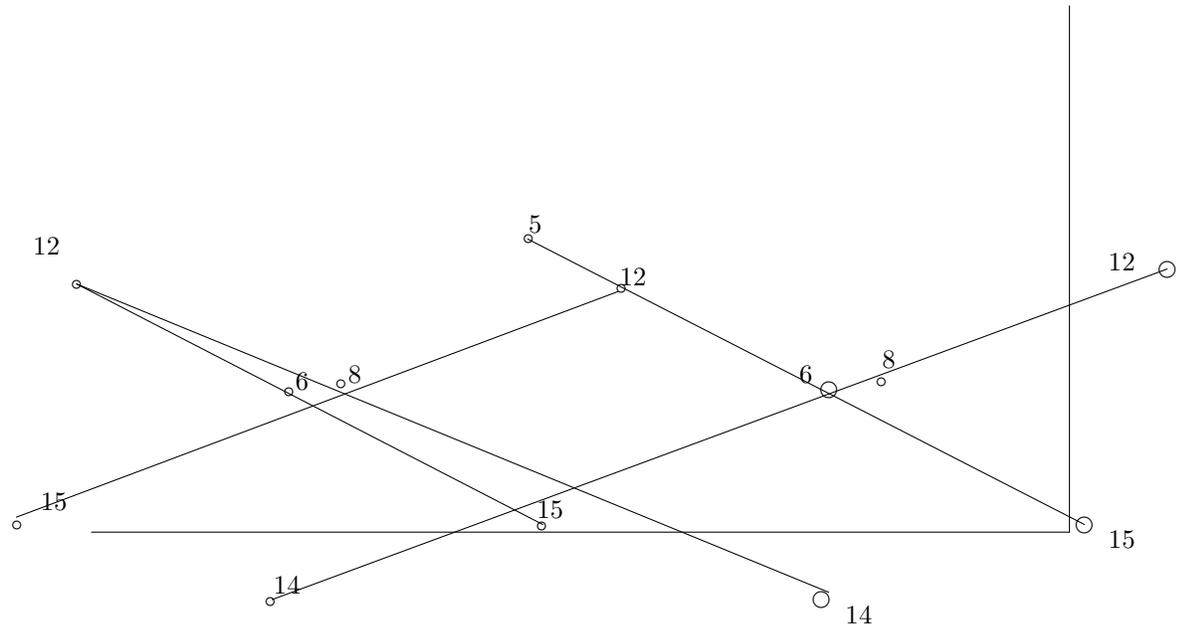
 5 $\alpha Cen$, 6 $\alpha Aur$, 8 $\alpha Cmi$,
 12 $\alpha Aql$, 14 $\alpha Tau$, 15 $\alpha Sco$
 on ($\lambda, \beta$) coordinate
 plane ( or on cilinder) -- Figure 10.

{\ }

It is an approximate  example of graph of a new kind,
 similar for graceful graphs $[ 12 ]$ and harmonious graphs $[ 13 ]$,
 where arithmetic, geometry and topology are tightly coupled.

 We can give definition:

{\ }

 \it{ each point i belongs to a straight line drawn through
 other two points j,k from the list, the sum of j and k being
 equal to 20}\rm.

 where is exception to the rule for point 5.

{\ }

  Also we have on the ($l,b$) and ($\alpha, \delta$ ) planes, Figures 11, 12 ---

{\ }

  \it{ for every points i,k,j which belong to a one of the straight lines
  there is other straight line with points i, j, k, or i, j, 20-k
  or i,j-20,k-20 (on ($l,b$)) }\rm.

{\ }

 Note that these are really the nearest extremally massive objects
 ( with exception of point 15 ).
 Moreover, there are the correlations with velocity vectors for
 presented configurations (or quasi-configurations) $[ 14 ]$ in the all
 planes by time.

{\ }

 Also we can construct similar figures for other viewpoints-
 view from $\alpha Cmi$ give to us Figure 13,
 view from $\alpha Cen$ Figure 14, view from $\alpha Aql$ Figure 15.
 We can see so there are same numbers--- 5, 6, 8, 12, 14, 15.

{\ }

 Let us consider the stars with the numbers $5k$ and  stars nearest
 by same order.
 We have paradoxical meshes of the parallel lines for these  stars--
 Fig. 16 ($\lambda, \beta$) plane, Figure 17 ($\alpha, \delta$ ) plane,
 and Figure 18-- for ($l,b$) planes (with parallelism on projective sense).
 Note that these are really extremally massive objects--
 more than 5 Sol's mass.

{\ }

 With rare exception, most part of $O,B$ and $A0$ stars (brighter
 than $6.0^m$) is concentrated near the straight lines drawn
 through pairs of the stars with numbers 2, 3,
 5, 6, 7, 10, 11, 15-- Figure 19; that is true
 for ($l,b$), ($\alpha,\delta$), ($\lambda,\beta$) coordinate planes.

{\ }

There is a substructure of stars of $\beta Per$ (EA) and $\beta Lyr$ (EB)
types,
 which concentrate only to some of the lines in all the coordinate systems.
 It is an independent test, the brightness of these stars is
 $8-9^m$ on the average. $K,M$ giants also gravitate towards these lines,
 but their positions are more smeared than those of $O,B$ stars.

{\ }

{\bf Conclusion \rm}

{\ }

 Perhaps  phenomenon does not completely visual and there are
order on native coordinate representation for equations (1) and (11).
Maybe gravitational interactions for far and large masses are masked
by heat motions of small stars and interstellar media.

{\ }

{\bf Reference \rm}

{\ }

1. Ambarcumjan V.A.  Proceedings of 2nd Cosmogony Conference Moscow, 1953.

2. Kholopoff P.N.  Communicatoins of GAISh 205, (1979), Moscow.

3. Ephremoff Ju. N.   Young Star Groups, in book :
   Stars and star systems, Science, Moscow 1981

4. Kopylov, A. I., Kuznetsov D. Y., Fetisova T. S., and Shvarzman V.
   F.  in Large Scale Structure of the Universe, eds. J. Audouze, M.--C.
   Pelletan, A. Szalay (Kluwer), (1988) 129 -- 137
   Nature, 385, 139 - 141, 9 January 1997

5. Einasto J.   A 120--MPC periodicity in the three-dimensional
   distribution of galaxy superclusters, Nature 385, (1997) 139 - 141

6. Hagihara Y.  Celestial Mechanics, Cambridge-Tokyo 1974,1975

7. DuboshinG.N. Celestial Mechanics, problems and methods, Science , Moscow,
1975

8.  Fix, G.F, and Strang, G : Theory of finite element methods, Mir, Moscow,
  1977

9.  Bellman R.  Introduction to matrix analisys
 New York, 1960

10. Gilbert E.H., Pollak H.O. Steiner minimal trees,
SIAM Journal of Applied Mathematics, 16,  1 (1968), 1-29

11. Kulikovski P.G.  Stars astronomy, Science,  Moscow 1985

12. Golomb S.W.  The Largest Graceful Subgraphof the
  Complete Graph, The American Mathematical Monthly 81, (1974),
  499-501.

13. Graham R., Sloane N.  On Additive Bases and Harmonious Graphs,
 SIAM Journal on Algebraic and Discrete Meth. 1,
 (1980),  382-404.

14. Levi F.  Geometrische Konfigurationenen, Lpz, 1929

\newpage
\unitlength=1mm
\special{em:linewidth 0.4pt}
\linethickness{0.4pt}
\begin{picture}(110.00,133.00)
\put(82.00,91.00){\circle{2.83}}
\put(97.00,113.00){\circle{2.83}}
\put(109.00,95.00){\circle{2.00}}
\put(94.00,132.00){\circle{2.00}}
\emline{83.00}{92.00}{1}{96.00}{112.00}{2}
\emline{98.00}{112.00}{3}{108.00}{96.00}{4}
\emline{94.00}{131.00}{5}{97.00}{114.00}{6}
\put(83.00,113.00){\makebox(0,0)[cc]{mass point $i$}}
\put(85.00,104.00){\makebox(0,0)[cc]{$R_{ij}$}}
\put(85.00,86.00){\makebox(0,0)[cc]{mass point $j$}}
\put(35.00,71.00){\vector(1,0){48.00}}
\put(35.00,71.00){\vector(0,1){36.00}}
\emline{35.00}{96.00}{7}{33.00}{96.00}{8}
\put(29.00,96.00){\makebox(0,0)[cc]{1.}}
\put(35.00,64.00){\makebox(0,0)[cc]{$i$}}
\put(52.00,64.00){\makebox(0,0)[cc]{$R_{ij}$}}
\emline{65.00}{71.00}{9}{65.00}{69.00}{10}
\put(65.00,65.00){\makebox(0,0)[cc]{$j$}}
\emline{35.00}{96.00}{11}{65.00}{71.00}{12}
\put(52.00,95.00){\makebox(0,0)[cc]{$v_i=v_i(l),$}}
\put(58.00,90.00){\makebox(0,0)[cc]{$v_{il}=1/R_{ij}$}}
\put(81.00,62.00){\makebox(0,0)[cc]{$l$}}
\put(30.00,106.00){\makebox(0,0)[cc]{$v$}}
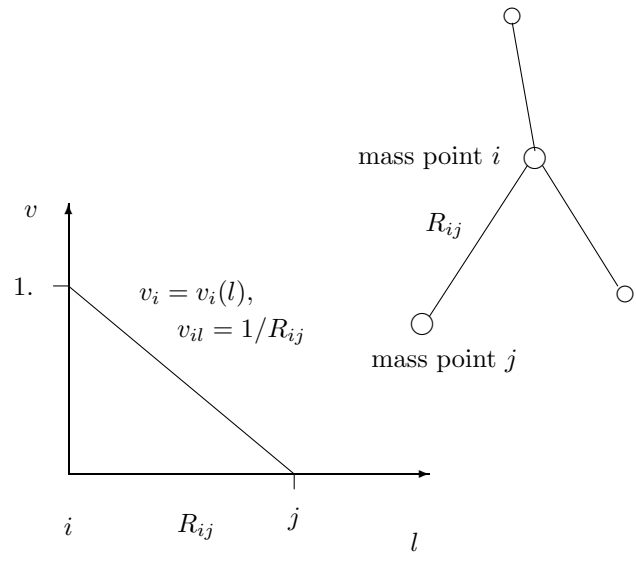
\put(70.00,50.00){\makebox(0,0)[cc]{Fig.1 Example of nodes related linear function}}
\end{picture}

\unitlength=1.00mm
\special{em:linewidth 0.4pt}
\linethickness{0.4pt}
\begin{picture}(140.00,140.00)
\emline{10.00}{70.00}{1}{140.00}{70.00}{2}
\emline{60.00}{70.00}{3}{60.00}{69.00}{4}
\emline{131.00}{70.00}{5}{131.00}{69.00}{6}
\put(59.00,63.00){\makebox(0,0)[cc]{0}}
\put(130.00,62.00){\makebox(0,0)[cc]{360}}
\emline{48.00}{70.00}{7}{48.00}{69.00}{8}
\put(45.00,62.00){\makebox(0,0)[cc]{300}}
\emline{10.00}{70.00}{9}{10.00}{69.00}{10}
\put(7.00,62.00){\makebox(0,0)[cc]{100}}
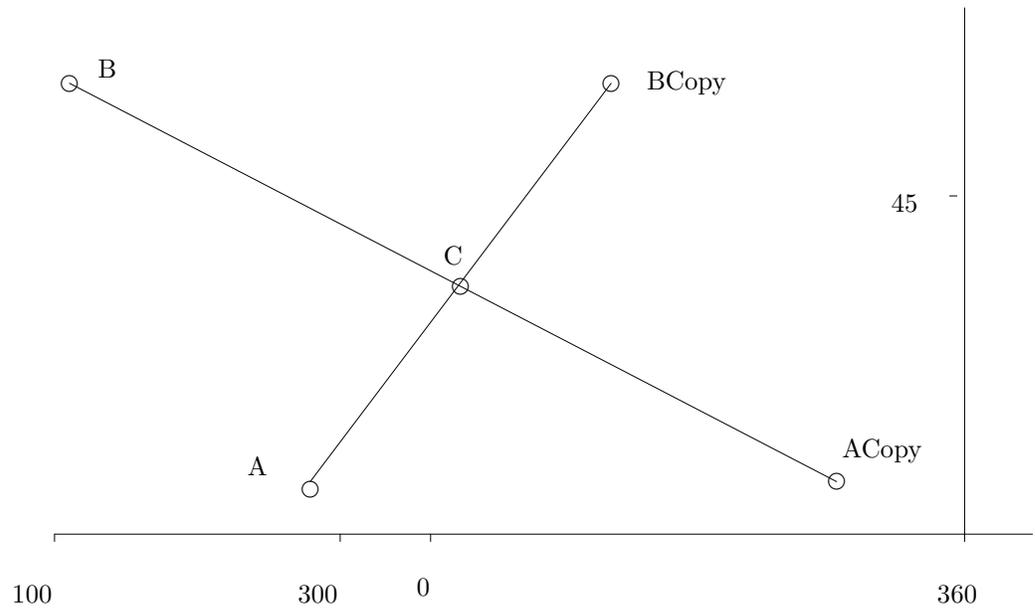
\put(63.00,28.00){\makebox(0,0)[cc]{Fig. 2.  Simulation, 4 points, spherical plane, one on the viewpoint.}}
\put(84.00,130.00){\circle{2.00}}
\put(114.00,77.00){\circle{2.00}}
\emline{44.00}{77.00}{11}{84.00}{130.00}{12}
\put(12.00,130.00){\circle{2.00}}
\emline{12.00}{130.00}{13}{114.00}{77.00}{14}
\put(44.00,76.00){\circle{2.00}}
\put(64.00,103.00){\circle{2.00}}
\put(37.00,79.00){\makebox(0,0)[cc]{A}}
\put(120.00,81.00){\makebox(0,0)[cc]{ACopy}}
\put(17.00,132.00){\makebox(0,0)[cc]{B}}
\put(94.00,130.00){\makebox(0,0)[cc]{BCopy}}
\put(63.00,107.00){\makebox(0,0)[cc]{C}}
\emline{131.00}{70.00}{15}{131.00}{140.00}{16}
\emline{130.00}{115.00}{17}{129.00}{115.00}{18}
\put(123.00,114.00){\makebox(0,0)[cc]{45}}
\end{picture}

\unitlength=1mm
\special{em:linewidth 0.4pt}
\linethickness{0.4pt}
\begin{picture}(155.00,91.00)
\emline{10.00}{70.00}{1}{140.00}{70.00}{2}
\emline{60.00}{70.00}{3}{60.00}{69.00}{4}
\emline{131.00}{70.00}{5}{131.00}{69.00}{6}
\put(59.00,63.00){\makebox(0,0)[cc]{0}}
\put(130.00,62.00){\makebox(0,0)[cc]{360}}
\emline{48.00}{70.00}{7}{48.00}{69.00}{8}
\put(45.00,62.00){\makebox(0,0)[cc]{300}}
\put(64.00,74.00){\circle{2.00}}
\put(101.00,50.00){\circle{2.00}}
\put(28.00,50.00){\circle{2.00}}
\emline{101.00}{50.00}{9}{38.00}{91.00}{10}
\put(83.00,61.00){\circle{2.00}}
\put(135.00,74.00){\circle{2.00}}
\emline{28.00}{50.00}{11}{135.00}{74.00}{12}
\put(63.00,78.00){\makebox(0,0)[cc]{A}}
\put(28.00,42.00){\makebox(0,0)[cc]{C}}
\put(104.00,42.00){\makebox(0,0)[cc]{C}}
\put(136.00,77.00){\makebox(0,0)[cc]{A}}
\emline{10.00}{70.00}{13}{10.00}{69.00}{14}
\put(7.00,62.00){\makebox(0,0)[cc]{100}}
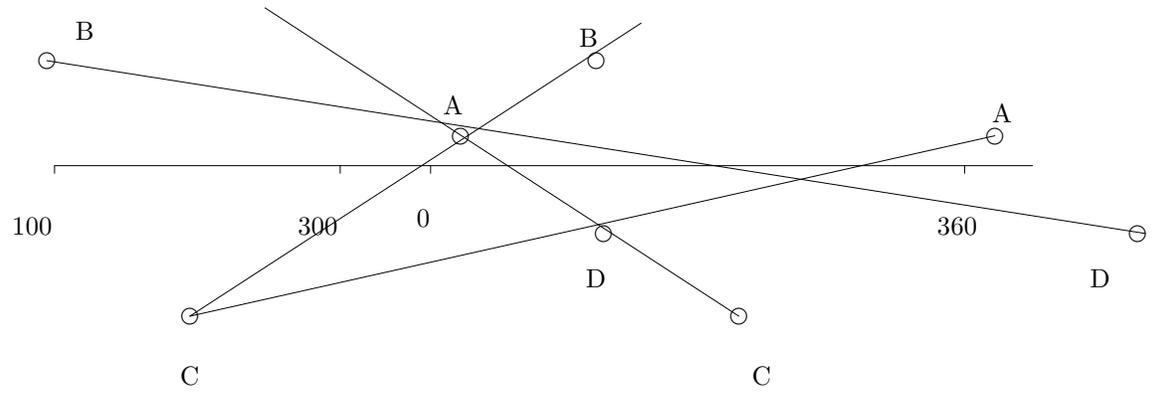
\put(63.00,18.00){\makebox(0,0)[cc]{Fig. 3. Simulations result, 5 points, one on the viewpoint.}}
\put(82.00,84.00){\circle{2.00}}
\emline{28.00}{50.00}{15}{88.00}{89.00}{16}
\put(81.00,87.00){\makebox(0,0)[cc]{B}}
\put(9.00,84.00){\circle{2.00}}
\put(14.00,88.00){\makebox(0,0)[cc]{B}}
\put(154.00,61.00){\circle{2.00}}
\emline{9.00}{84.00}{17}{155.00}{61.00}{18}
\put(82.00,55.00){\makebox(0,0)[cc]{D}}
\put(149.00,55.00){\makebox(0,0)[cc]{D}}
\end{picture}

\newpage

\unitlength=1mm
\special{em:linewidth 0.4pt}
\linethickness{0.4pt}
\begin{picture}(146.00,151.00)
\emline{10.00}{70.00}{1}{140.00}{70.00}{2}
\emline{140.00}{70.00}{3}{140.00}{140.00}{4}
\put(35.00,99.00){\makebox(0,0)[cc]{Fig. 4. Simulation result, some points.}}
\emline{60.00}{70.00}{5}{60.00}{69.00}{6}
\put(60.00,64.00){\makebox(0,0)[cc]{0}}
\put(109.00,71.00){\circle{2.00}}
\put(36.00,29.00){\circle{2.00}}
\put(19.00,150.00){\circle{2.00}}
\emline{19.00}{150.00}{7}{146.00}{39.00}{8}
\put(52.00,12.00){\circle{2.00}}
\emline{52.00}{11.00}{9}{52.00}{12.00}{10}
\emline{52.00}{12.00}{11}{7.00}{60.00}{12}
\put(107.00,29.00){\circle{2.00}}
\put(98.00,6.00){\circle{2.00}}
\emline{98.00}{6.00}{13}{136.00}{105.00}{14}
\put(91.00,150.00){\circle{2.00}}
\put(25.00,6.00){\circle{2.00}}
\emline{25.00}{6.00}{15}{91.00}{150.00}{16}
\emline{91.00}{150.00}{17}{121.00}{18.00}{18}
\put(121.00,18.00){\circle{2.00}}
\put(48.00,18.00){\circle{2.00}}
\put(70.00,106.00){\circle{2.00}}
\emline{140.00}{115.00}{19}{138.00}{115.00}{20}
\put(134.00,118.00){\makebox(0,0)[cc]{45}}
\put(137.00,106.00){\circle{2.00}}
\emline{137.00}{106.00}{21}{53.00}{11.00}{22}
\end{picture}

\unitlength=1mm
\special{em:linewidth 0.4pt}
\linethickness{0.4pt}
\begin{picture}(121.00,131.41)
\put(50.00,120.00){\circle{2.00}}
\put(50.00,100.00){\circle{2.00}}
\put(50.00,80.00){\circle{2.00}}
\put(75.00,110.00){\circle{2.00}}
\put(75.00,90.00){\circle{2.00}}
\put(75.00,70.00){\circle{2.00}}
\put(58.00,60.00){\circle{2.00}}
\put(67.00,130.00){\circle{2.00}}
\put(60.00,80.00){\circle*{2.00}}
\put(64.00,90.00){\circle*{2.00}}
\put(60.00,100.00){\circle*{2.00}}
\put(64.00,110.00){\circle*{2.00}}
\put(60.00,120.00){\circle*{2.00}}
\put(64.00,70.00){\circle*{2.00}}
\emline{64.00}{70.00}{1}{59.00}{61.00}{2}
\emline{64.00}{70.00}{3}{75.00}{70.00}{4}
\emline{64.00}{70.00}{5}{60.00}{80.00}{6}
\emline{60.00}{80.00}{7}{51.00}{80.00}{8}
\emline{60.00}{80.00}{9}{64.00}{90.00}{10}
\emline{64.00}{90.00}{11}{74.00}{90.00}{12}
\emline{60.00}{100.00}{13}{64.00}{90.00}{14}
\emline{60.00}{100.00}{15}{51.00}{100.00}{16}
\emline{60.00}{100.00}{17}{64.00}{110.00}{18}
\emline{64.00}{110.00}{19}{75.00}{110.00}{20}
\emline{64.00}{110.00}{21}{60.00}{120.00}{22}
\emline{60.00}{120.00}{23}{67.00}{129.00}{24}
\emline{60.00}{120.00}{25}{51.00}{120.00}{26}
\emline{82.00}{95.00}{27}{120.00}{95.00}{28}
\emline{120.00}{95.00}{29}{100.00}{130.00}{30}
\emline{100.00}{130.00}{31}{83.00}{95.00}{32}
\emline{83.00}{95.00}{33}{95.00}{81.00}{34}
\emline{95.00}{81.00}{35}{120.00}{95.00}{36}
\emline{95.00}{81.00}{37}{100.00}{130.00}{38}
\put(100.00,130.00){\circle{2.83}}
\put(120.00,95.00){\circle{2.00}}
\put(83.00,95.00){\circle{2.00}}
\put(95.00,81.00){\circle{2.00}}
\put(101.00,93.00){\circle*{2.00}}
\put(95.00,101.00){\circle*{2.00}}
\put(101.00,93.00){\vector(-1,-2){4.00}}
\put(102.00,93.00){\vector(-3,4){4.67}}
\put(97.33,99.00){\vector(0,0){0.00}}
\put(102.00,93.00){\vector(1,0){10.00}}
\put(59.00,33.00){\makebox(0,0)[cc]{Fig.5 Circles - 'boundary' points, filled circles-points }}
\put(59.00,28.00){\makebox(0,0)[cc]{ in the 'volume', 2 and 3 D Steiner configurations,}}

\emline{22.00}{99.00}{39}{37.00}{117.00}{40}
\emline{37.00}{117.00}{41}{46.00}{92.00}{42}
\emline{46.00}{92.00}{43}{22.00}{99.00}{44}
\put(37.00,117.00){\circle{2.00}}
\put(23.00,99.00){\circle{2.00}}
\put(47.00,92.00){\circle{2.00}}
\put(40.00,98.00){\circle*{2.00}}
\put(40.00,98.00){\vector(-4,3){4.00}}
\end{picture}

\newpage
\unitlength=1mm
\special{em:linewidth 0.4pt}
\linethickness{0.4pt}
\begin{picture}(143.33,134.00)
\put(72.33,70.00){\line(1,0){66.00}}
\put(72.67,70.00){\line(0,1){60.00}}
\put(72.00,134.00){\makebox(0,0)[cc]{b}}
\put(143.33,70.33){\makebox(0,0)[cc]{l}}
\put(65.33,31.33){\makebox(0,0)[cc]{ Fig. 6. Light and nearest galaxies, lighter than $8.5^m$ -(l,b) representation.}}
\put(100.00,70.00){\circle{1.49}}
\put(103.67,66.00){\makebox(0,0)[cc]{Dwingeloo1 }}
\put(29.00,70.00){\circle{1.49}}
\put(11.33,71.67){\makebox(0,0)[cc]{Dwingeloo1 }}
\put(96.00,50.00){\circle{1.33}}
\put(103.00,48.33){\makebox(0,0)[cc]{M31}}
\put(24.00,50.00){\circle{1.49}}
\put(18.67,52.67){\makebox(0,0)[cc]{M31}}
\put(62.33,88.33){\circle{1.49}}
\put(61.00,91.00){\makebox(0,0)[cc]{NGC5128}}
\put(99.67,110.00){\circle{1.49}}
\emline{99.67}{110.00}{1}{28.67}{69.67}{2}
\put(99.67,113.67){\makebox(0,0)[cc]{M81}}
\put(92.00,129.00){\circle{1.49}}
\put(95.00,132.33){\makebox(0,0)[cc]{M101}}
\put(27.33,40.00){\circle{1.49}}
\emline{27.33}{40.00}{3}{92.33}{129.00}{4}
\put(33.67,37.67){\makebox(0,0)[cc]{M33}}
\put(20.33,128.67){\circle{1.49}}
\emline{20.67}{128.67}{5}{96.33}{49.67}{6}
\put(25.33,129.00){\makebox(0,0)[cc]{M101}}
\put(61.33,104.00){\makebox(0,0)[cc]{M83}}
\put(61.67,100.00){\circle{1.49}}
\emline{29.33}{70.33}{7}{92.33}{128.67}{8}
\put(29.00,110.33){\circle{1.33}}
\emline{29.00}{110.33}{9}{100.67}{69.67}{10}
\end{picture}

\newpage
\unitlength=1.00mm
\special{em:linewidth 0.4pt}
\linethickness{0.4pt}
\begin{picture}(154.00,143.00)
\emline{72.00}{70.00}{1}{137.00}{70.00}{2}
\emline{72.00}{70.00}{3}{72.00}{140.00}{4}
\put(72.00,143.00){\makebox(0,0)[cc]{$\beta$}}
\put(140.00,70.00){\makebox(0,0)[cc]{$\lambda$}}
\put(81.00,89.00){\circle{2.00}}
\put(81.00,91.00){\makebox(0,0)[cc]{M33}}
\put(108.00,130.00){\circle{2.00}}
\put(108.00,132.00){\makebox(0,0)[cc]{M101}}
\put(44.00,47.00){\makebox(0,0)[cc]{NGC5128}}
\put(43.00,39.00){\circle{0.00}}
\put(43.00,39.00){\circle{2.00}}
\emline{43.00}{39.00}{5}{108.00}{130.00}{6}
\put(36.00,130.00){\circle{2.00}}
\put(36.00,132.00){\makebox(0,0)[cc]{M101}}
\put(115.00,39.00){\circle{2.00}}
\put(114.00,52.00){\circle{2.00}}
\emline{115.00}{39.00}{7}{108.00}{130.00}{8}
\emline{114.00}{52.00}{9}{36.00}{130.00}{10}
\put(120.00,51.00){\makebox(0,0)[cc]{M83}}
\put(84.00,110.00){\circle{2.00}}
\put(94.00,119.00){\circle{2.00}}
\emline{108.00}{130.00}{11}{44.00}{78.00}{12}
\put(93.00,122.00){\makebox(0,0)[cc]{M81}}
\put(76.00,120.00){\makebox(0,0)[cc]{Dwingeloo 1}}
\put(77.00,104.00){\circle{2.00}}
\put(77.00,99.00){\makebox(0,0)[cc]{M31}}
\put(49.00,11.00){\makebox(0,0)[cc]{Fig. 7.  Light and nearest galaxies, lighter than $8.5^m$,   }}
\put(49.00,6.00){\makebox(0,0)[cc]{  $(\lambda, \beta)$ representation. }}
\put(42.00,52.00){\circle{2.00}}
\put(42.00,55.00){\makebox(0,0)[cc]{M83}}
\put(122.00,34.00){\makebox(0,0)[cc]{NGC5128}}
\put(153.00,89.00){\circle{2.00}}
\emline{152.00}{90.00}{13}{77.00}{104.00}{14}
\emline{115.00}{39.00}{15}{17.00}{56.00}{16}
\put(146.00,83.00){\makebox(0,0)[cc]{M33}}
\end{picture}

\newpage
\unitlength=1mm
\special{em:linewidth 0.4pt}
\linethickness{0.4pt}
\begin{picture}(144.00,142.00)
\emline{10.00}{70.00}{1}{144.00}{70.00}{2}
\put(84.00,129.00){\circle{2.00}}
\put(86.00,134.00){\makebox(0,0)[cc]{Dw1}}
\put(77.00,111.00){\circle{2.00}}
\put(83.00,111.00){\makebox(0,0)[cc]{M31}}
\put(81.00,100.00){\circle{2.00}}
\put(87.00,99.00){\makebox(0,0)[cc]{M33}}
\put(115.00,25.00){\circle{2.00}}
\put(125.00,25.00){\makebox(0,0)[cc]{Ngc 5128}}
\put(40.00,25.00){\circle{2.00}}
\emline{84.00}{129.00}{3}{40.00}{25.00}{4}
\put(104.00,139.00){\circle{2.00}}
\put(9.00,100.00){\circle{2.00}}
\emline{9.00}{100.00}{5}{104.00}{139.00}{6}
\put(113.00,139.00){\makebox(0,0)[cc]{M81,82}}
\put(117.00,124.00){\circle{2.00}}
\put(32.00,139.00){\circle{2.00}}
\emline{32.00}{139.00}{7}{118.00}{123.00}{8}
\put(126.00,123.00){\makebox(0,0)[cc]{M101}}
\put(42.00,142.00){\makebox(0,0)[cc]{M81,82}}
\emline{33.00}{138.00}{9}{115.00}{25.00}{10}
\put(44.00,124.00){\circle{2.00}}
\put(49.00,126.00){\makebox(0,0)[cc]{M101}}
\emline{117.00}{124.00}{11}{14.00}{56.00}{12}
\emline{14.00}{56.00}{13}{17.00}{60.00}{14}
\emline{17.00}{60.00}{15}{19.00}{55.00}{16}
\emline{19.00}{55.00}{17}{14.00}{56.00}{18}
\put(32.00,56.00){\makebox(0,0)[cc]{to Ngc 5128}}
\put(13.00,94.00){\makebox(0,0)[cc]{M33}}
\put(53.00,21.00){\makebox(0,0)[cc]{Ngc 5128}}
\put(72.00,-7.00){\makebox(0,0)[cc]{Fig. 8.  $(\alpha, \delta)$- representation. }}
\emline{75.00}{70.00}{19}{75.00}{67.00}{20}
\emline{111.00}{70.00}{21}{111.00}{66.00}{22}
\put(75.00,61.00){\makebox(0,0)[cc]{0}}
\put(110.00,59.00){\makebox(0,0)[cc]{$\pi$}}
\emline{39.00}{70.00}{23}{39.00}{66.00}{24}
\put(35.00,61.00){\makebox(0,0)[cc]{$-\pi$}}
\end{picture}
\newpage
\unitlength=1mm
\special{em:linewidth 0.4pt}
\linethickness{0.4pt}
\begin{picture}(148.00,95.00)
\emline{10.00}{70.00}{1}{148.00}{70.00}{2}
\put(88.00,93.00){\circle{2.00}}
\put(107.00,65.00){\circle{2.00}}
\put(18.00,93.00){\circle{2.00}}
\put(83.00,71.00){\circle{2.00}}
\emline{107.00}{65.00}{3}{18.00}{93.00}{4}
\put(83.00,72.00){\circle{2.00}}
\put(70.00,82.00){\circle{2.00}}
\put(38.00,65.00){\circle{2.00}}
\emline{38.00}{65.00}{5}{88.00}{93.00}{6}
\put(70.00,70.00){\circle{2.00}}
\put(0.00,82.00){\circle{2.00}}
\emline{0.00}{82.00}{7}{107.00}{65.00}{8}
\put(75.00,69.00){\circle{2.00}}
\put(92.00,94.00){\makebox(0,0)[cc]{M101}}
\put(20.00,95.00){\makebox(0,0)[cc]{M101}}
\put(107.00,59.00){\makebox(0,0)[cc]{Ngc5128}}
\put(37.00,58.00){\makebox(0,0)[cc]{Ngc5128}}
\put(68.00,85.00){\makebox(0,0)[cc]{M31}}
\put(11.00,84.00){\makebox(0,0)[cc]{M31}}
\put(86.00,74.00){\makebox(0,0)[cc]{M81,82}}
\put(75.00,63.00){\makebox(0,0)[cc]{Dw1}}
\put(64.00,66.00){\makebox(0,0)[cc]{M33}}
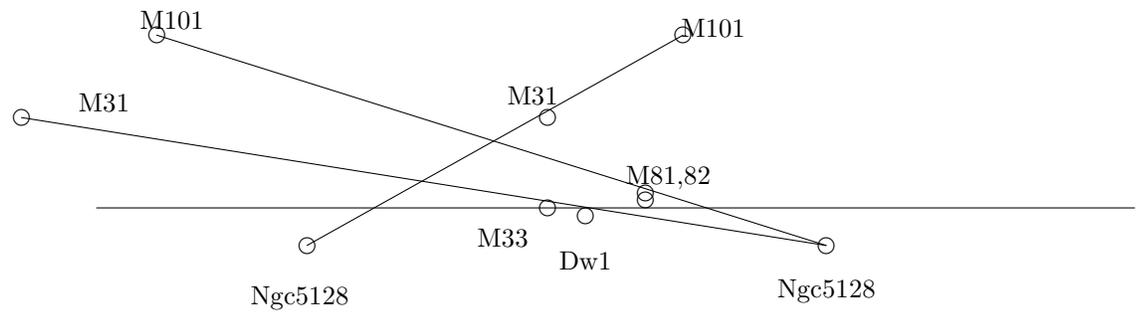
\put(60.00,40.00){\makebox(0,0)[cc]{Fig. 9. Supergalactic plane representation.}}
\end{picture}

\newpage
\unitlength=1.00mm
\special{em:linewidth 0.4pt}
\linethickness{0.4pt}
\begin{picture}(154.00,140.00)
\emline{140.00}{70.00}{1}{140.00}{140.00}{2}
\emline{140.00}{70.00}{3}{10.00}{70.00}{4}
\put(69.00,111.00){\makebox(0,0)[cc]{5}}
\
\put(68.08,109.00){\circle{1.00}}
\put(38.00,90.00){\makebox(0,0)[cc]{6}}
\
\put(36.27,88.64){\circle{1.00}}
\put(45.00,91.00){\makebox(0,0)[cc]{8}}
\
\put(43.15,89.66){\circle{1.00}}
\put(82.00,104.00){\makebox(0,0)[cc]{12}}
\
\put(80.42,102.36){\circle{1.00}}
\put(36.00,63.00){\makebox(0,0)[cc]{14}}
\
\put(33.77,60.74){\circle{1.00}}
\put(71.00,73.00){\makebox(0,0)[cc]{15}}
\
\put(69.81,70.81){\circle{1.00}}
\put(43.00,15.00){\makebox(0,0)[cc]{Fig 10. 5- $\alpha Cen$, 6-$\alpha Aur$,}}
\put(43.00,7.00){\makebox(0,0)[cc]{ 8- $alpha Cmi$, 12 -$\alpha Aql$, 14- $\alpha Tau$, 15- $\alpha Sco$ ($\lambda, \beta$)
 plane. }}
\
\put(8.00,103.00){\circle{1.00}}
\put(4.00,108.00){\makebox(0,0)[cc]{12}}
\put(0.10,71.00){\circle{1.00}}
\put(5.00,74.00){\makebox(0,0)[cc]{15}}
\put(115.00,90.00){\circle{1.00}}
\put(116.00,93.00){\makebox(0,0)[cc]{8}}
\emline{0.00}{72.00}{5}{80.00}{102.00}{6}
\emline{8.00}{103.00}{7}{70.00}{71.00}{8}
\put(153.00,105.00){\circle{2.00}}
\emline{34.00}{61.00}{9}{153.00}{105.00}{10}
\put(147.00,106.00){\makebox(0,0)[cc]{12}}
\put(142.00,71.00){\circle{2.00}}
\emline{68.00}{109.00}{11}{142.00}{71.00}{12}
\put(107.00,61.00){\circle{2.00}}
\emline{8.00}{103.00}{13}{108.00}{62.00}{14}
\put(112.00,59.00){\makebox(0,0)[cc]{14}}
\put(147.00,69.00){\makebox(0,0)[cc]{15}}
\put(108.00,89.00){\circle{2.00}}
\put(105.00,91.00){\makebox(0,0)[cc]{6}}
\end{picture}
\newpage
\unitlength=1.00mm
\special{em:linewidth 0.4pt}
\linethickness{0.4pt}
\begin{picture}(151.00,140.00)
\emline{20.00}{80.00}{1}{140.00}{80.00}{2}
\emline{140.00}{80.00}{3}{140.00}{140.00}{4}
\put(40.00,62.00){\circle{2.00}}
\put(93.00,82.00){\circle{2.00}}
\put(145.00,106.00){\circle{2.00}}
\put(112.00,62.00){\circle{2.00}}
\put(73.00,106.00){\circle{2.00}}
\emline{73.00}{106.00}{5}{112.00}{62.00}{6}
\put(73.00,109.00){\makebox(0,0)[cc]{8}}
\put(114.00,57.00){\makebox(0,0)[cc]{12}}
\put(91.00,76.00){\makebox(0,0)[cc]{5}}
\put(33.00,56.00){\makebox(0,0)[cc]{12}}
\put(145.00,110.00){\makebox(0,0)[cc]{8}}
\put(135.00,131.00){\makebox(0,0)[cc]{b}}
\put(23.00,73.00){\makebox(0,0)[cc]{l}}
\put(47.00,40.00){\makebox(0,0)[cc]{Fig. 11 (l,b) plane.}}
\emline{40.00}{62.00}{7}{73.00}{106.00}{8}
\put(63.00,91.00){\circle{2.00}}
\emline{113.00}{62.00}{9}{25.00}{113.00}{10}
\emline{25.00}{113.00}{11}{139.00}{47.00}{12}
\put(68.00,91.00){\makebox(0,0)[cc]{6}}
\
\put(138.00,40.00){\circle{2.00}}
\put(30.00,110.00){\circle{2.00}}
\put(32.00,114.00){\makebox(0,0)[cc]{15}}
\put(1.00,40.00){\circle{2.00}}
\emline{145.00}{106.00}{13}{0.00}{43.00}{14}
\put(6.00,38.00){\makebox(0,0)[cc]{14}}
\put(133.00,41.00){\makebox(0,0)[cc]{14}}
\emline{145.00}{106.00}{15}{0.00}{80.00}{16}
\emline{0.00}{80.00}{17}{2.00}{80.00}{18}
\put(20.00,82.00){\circle{2.00}}
\put(18.00,84.00){\makebox(0,0)[cc]{5}}
\emline{112.00}{62.00}{19}{145.00}{106.00}{20}
\put(134.00,91.00){\circle{0.00}}
\put(134.00,91.00){\circle{0.00}}
\put(135.00,91.00){\circle{0.00}}
\put(135.00,91.00){\circle{2.00}}
\put(137.00,87.00){\makebox(0,0)[cc]{6}}
\emline{75.00}{105.00}{21}{151.00}{87.00}{22}
\put(145.00,85.00){\makebox(0,0)[cc]{to 5}}
\end{picture}

\newpage
\unitlength=1.00mm
\special{em:linewidth 0.4pt}
\linethickness{0.4pt}
\begin{picture}(144.00,140.00)
\emline{140.00}{70.00}{1}{140.00}{140.00}{2}
\
\
\
\
\
\
\
\emline{140.00}{70.00}{3}{10.00}{70.00}{4}
\put(5.00,44.00){\circle{2.00}}
\put(77.00,44.00){\circle{2.00}}
\put(43.00,86.00){\circle{2.00}}
\put(52.00,75.00){\circle{2.00}}
\emline{43.00}{86.00}{5}{77.00}{44.00}{6}
\put(15.00,78.00){\circle{2.00}}
\put(87.00,79.00){\circle{2.00}}
\put(115.00,115.00){\circle{2.00}}
\emline{115.00}{115.00}{7}{5.00}{44.00}{8}
\put(143.00,10.00){\circle{2.00}}
\emline{143.00}{10.00}{9}{9.00}{81.00}{10}
\put(124.00,75.00){\circle{2.00}}
\emline{43.00}{86.00}{11}{124.00}{75.00}{12}
\emline{143.00}{9.00}{13}{116.00}{115.00}{14}
\put(15.00,10.00){\circle{2.00}}
\emline{15.00}{10.00}{15}{124.00}{75.00}{16}
\put(86.00,75.00){\makebox(0,0)[cc]{12}}
\put(44.00,88.00){\makebox(0,0)[cc]{14}}
\put(112.00,118.00){\makebox(0,0)[cc]{6}}
\put(128.00,76.00){\makebox(0,0)[cc]{8}}
\put(57.00,74.00){\makebox(0,0)[cc]{8}}
\put(15.00,81.00){\makebox(0,0)[cc]{12}}
\put(74.00,37.00){\makebox(0,0)[cc]{15}}
\put(13.00,14.00){\makebox(0,0)[cc]{5}}
\put(135.00,9.00){\makebox(0,0)[cc]{5}}
\put(12.00,40.00){\makebox(0,0)[cc]{15}}
\put(82.00,27.00){\makebox(0,0)[cc]{Fig 12 ($\alpha , \delta$) plane.}}
\end{picture}
\newpage
\unitlength=1mm
\special{em:linewidth 0.4pt}
\linethickness{0.4pt}
\begin{picture}(140.00,142.00)
\emline{0.00}{70.00}{1}{140.00}{70.00}{2}
\emline{20.00}{70.00}{3}{20.00}{142.00}{4}
\put(22.00,121.00){\circle{0.00}}
\put(22.00,121.00){\circle{2.00}}
\put(38.00,97.00){\circle{2.00}}
\put(53.00,69.00){\circle{2.00}}
\put(59.00,120.00){\circle{0.00}}
\put(58.00,121.00){\circle{2.83}}
\put(17.00,69.00){\circle{2.00}}
\emline{17.00}{69.00}{5}{64.00}{129.00}{6}
\put(75.00,97.00){\circle{2.00}}
\put(37.00,44.00){\circle{2.00}}
\emline{37.00}{44.00}{7}{90.00}{120.00}{8}
\put(74.00,44.00){\circle{2.00}}
\put(48.00,83.00){\circle{2.00}}
\emline{37.00}{44.00}{9}{61.00}{133.00}{10}
\emline{22.00}{121.00}{11}{75.00}{43.00}{12}
\put(24.00,122.00){\makebox(0,0)[cc]{5}}
\put(18.00,64.00){\makebox(0,0)[cc]{12}}
\put(54.00,63.00){\makebox(0,0)[cc]{12}}
\put(138.00,72.00){\makebox(0,0)[cc]{12}}
\put(40.00,40.00){\makebox(0,0)[cc]{8}}
\put(78.00,39.00){\makebox(0,0)[cc]{8}}
\put(62.00,118.00){\makebox(0,0)[cc]{5}}
\put(39.00,102.00){\makebox(0,0)[cc]{6}}
\put(53.00,81.00){\makebox(0,0)[cc]{15}}
\put(75.00,101.00){\makebox(0,0)[cc]{6}}
\put(75.00,90.00){\makebox(0,0)[cc]{14}}
\put(22.00,138.00){\makebox(0,0)[cc]{b}}
\put(114.00,62.00){\makebox(0,0)[cc]{l}}
\put(38.00,94.00){\circle{2.00}}
\put(37.00,88.00){\makebox(0,0)[cc]{14}}
\put(128.00,69.00){\circle{2.00}}
\put(75.00,94.00){\circle{2.00}}
\emline{22.00}{120.00}{13}{132.00}{67.00}{14}
\put(7.00,57.00){\makebox(0,0)[cc]{to 8}}
\put(130.00,121.00){\circle{2.00}}
\put(129.00,124.00){\makebox(0,0)[cc]{5}}
\emline{130.00}{121.00}{15}{4.00}{63.00}{16}
\emline{4.00}{63.00}{17}{4.00}{63.00}{18}
\emline{4.00}{63.00}{19}{7.00}{66.00}{20}
\emline{7.00}{66.00}{21}{9.00}{64.00}{22}
\emline{9.00}{64.00}{23}{4.00}{63.00}{24}
\put(47.00,22.00){\makebox(0,0)[cc]{Fig 13. Viewpoint is $\alpha Cmi$, 5-$\alpha Boo,$}}
\put(47.00,15.00){\makebox(0,0)[cc]{6- $\beta Gem$, 8-$\alpha Tau$, 12- $\alpha Cru$, 14- $\alpha Gem$, 15- $\alpha Sco$}}
\end{picture}

\newpage
\unitlength=1mm
\special{em:linewidth 0.4pt}
\linethickness{0.4pt}
\begin{picture}(146.00,130.00)
\emline{0.00}{70.00}{1}{146.00}{70.00}{2}
\emline{40.00}{70.00}{3}{40.00}{130.00}{4}
\put(55.00,87.00){\circle{2.00}}
\put(90.00,61.00){\circle{2.00}}
\put(31.00,122.00){\circle{2.00}}
\put(106.00,70.00){\circle{2.00}}
\put(34.00,70.00){\circle{2.00}}
\emline{31.00}{122.00}{5}{94.00}{57.00}{6}
\put(19.00,61.00){\circle{2.00}}
\emline{112.00}{70.00}{7}{112.00}{67.00}{8}
\put(103.00,122.00){\circle{2.00}}
\emline{103.00}{122.00}{9}{10.00}{55.00}{10}
\put(95.00,85.00){\circle{2.00}}
\emline{103.00}{122.00}{11}{90.00}{61.00}{12}
\put(23.00,85.00){\circle{2.00}}
\emline{23.00}{85.00}{13}{107.00}{70.00}{14}
\put(72.00,75.00){\circle{2.00}}
\emline{55.00}{86.00}{15}{91.00}{61.00}{16}
\put(27.00,124.00){\makebox(0,0)[cc]{15}}
\put(107.00,122.00){\makebox(0,0)[cc]{15}}
\put(99.00,84.00){\makebox(0,0)[cc]{14}}
\put(20.00,88.00){\makebox(0,0)[cc]{14}}
\put(68.00,71.00){\makebox(0,0)[cc]{6}}
\put(51.00,88.00){\makebox(0,0)[cc]{5}}
\put(108.00,72.00){\makebox(0,0)[cc]{12}}
\put(34.00,65.00){\makebox(0,0)[cc]{12}}
\put(22.00,54.00){\makebox(0,0)[cc]{8}}
\put(86.00,54.00){\makebox(0,0)[cc]{8}}
\put(140.00,64.00){\makebox(0,0)[cc]{l}}
\put(43.00,126.00){\makebox(0,0)[cc]{b}}
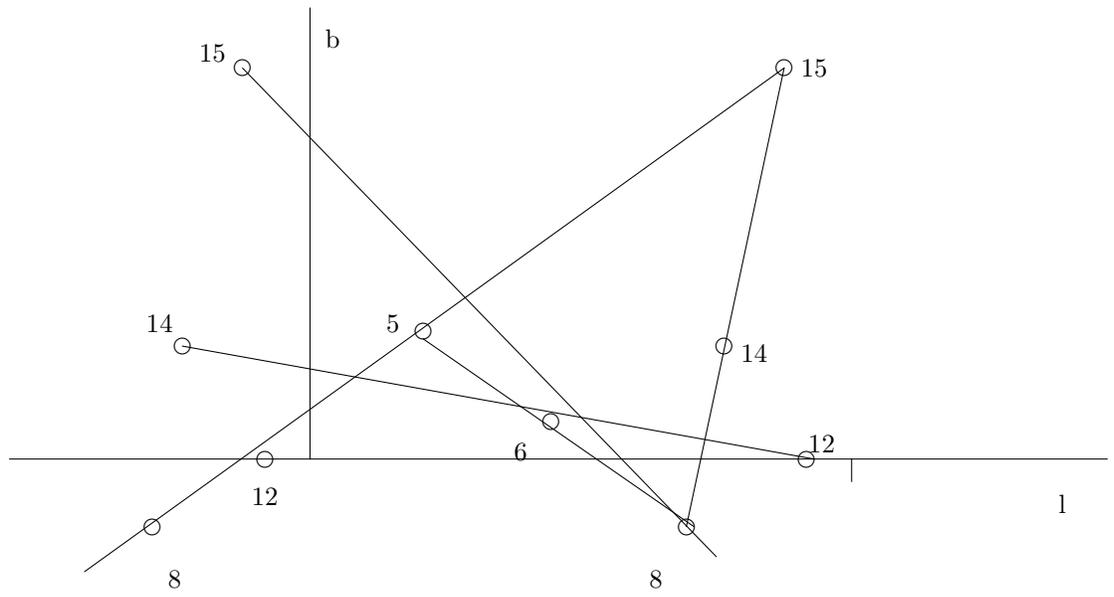
\put(40.00,40.00){\makebox(0,0)[cc]{Fig. 14 Viewpoint is $\alpha Cen$, 5-$\alpha Lyr$,6- $\alpha Aur$,8-$\alpha Ori$,}}
\put(39.00,31.00){\makebox(0,0)[cc]{12- $\alpha Cru$, 14- $\alpha Sco$, 15- $\alpha Vir$}}
\end{picture}

\unitlength=1.00mm
\special{em:linewidth 0.4pt}
\linethickness{0.4pt}
\begin{picture}(146.00,145.00)
\emline{20.00}{70.00}{1}{20.00}{145.00}{2}
\emline{20.00}{70.00}{3}{115.00}{70.00}{4}
\put(56.00,77.00){\circle{2.00}}
\put(12.00,68.00){\circle{2.00}}
\put(37.00,72.00){\circle{2.00}}
\emline{12.00}{68.00}{5}{56.00}{77.00}{6}
\put(72.00,55.00){\circle{2.00}}
\put(70.00,62.00){\circle{2.00}}
\put(84.00,68.00){\circle{0.00}}
\put(84.00,68.00){\circle{2.00}}
\put(110.00,72.00){\circle{2.00}}
\put(130.00,77.00){\circle{2.00}}
\emline{130.00}{77.00}{7}{68.00}{62.00}{8}
\put(144.00,55.00){\circle{2.00}}
\emline{56.00}{77.00}{9}{144.00}{55.00}{10}
\put(140.00,62.00){\circle{2.00}}
\emline{37.00}{72.00}{11}{140.00}{62.00}{12}
\emline{56.00}{77.00}{13}{72.00}{55.00}{14}
\put(65.00,73.00){\circle{0.00}}
\put(65.00,73.00){\circle{2.00}}
\emline{37.00}{72.00}{15}{130.00}{77.00}{16}
\put(56.00,79.00){\makebox(0,0)[cc]{5}}
\put(12.00,62.00){\makebox(0,0)[cc]{8}}
\put(34.00,74.00){\makebox(0,0)[cc]{15}}
\put(67.00,76.00){\makebox(0,0)[cc]{12}}
\put(146.00,49.00){\makebox(0,0)[cc]{14}}
\put(86.00,63.00){\makebox(0,0)[cc]{8}}
\put(74.00,49.00){\makebox(0,0)[cc]{14}}
\put(74.00,59.00){\makebox(0,0)[cc]{6}}
\put(110.00,78.00){\makebox(0,0)[cc]{15}}
\put(132.00,79.00){\makebox(0,0)[cc]{5}}
\put(141.00,64.00){\makebox(0,0)[cc]{6}}
\put(25.00,140.00){\makebox(0,0)[cc]{b}}
\put(39.00,33.00){\makebox(0,0)[cc]{Fig. 15. Viewpoint $\alpha Aql$}}
\end{picture}

\newpage
\unitlength=1.00mm
\special{em:linewidth 0.4pt}
\linethickness{0.4pt}
\begin{picture}(159.00,153.00)
\emline{10.00}{70.00}{1}{139.00}{70.00}{2}
\emline{139.00}{70.00}{3}{139.00}{140.00}{4}
\put(22.00,149.00){\circle{2.00}}
\put(68.00,131.00){\circle{2.00}}
\put(24.00,150.00){\makebox(0,0)[cc]{2}}
\put(69.00,134.00){\makebox(0,0)[cc]{20}}
\emline{22.00}{149.00}{5}{152.00}{98.00}{6}
\put(120.00,110.00){\circle{2.00}}
\put(122.00,112.00){\makebox(0,0)[cc]{5}}
\put(119.00,112.00){\circle{2.00}}
\put(116.00,114.00){\makebox(0,0)[cc]{11}}
\put(16.00,89.00){\circle{2.00}}
\put(18.00,91.00){\makebox(0,0)[cc]{6}}
\put(50.00,71.00){\circle{2.00}}
\put(52.00,74.00){\makebox(0,0)[cc]{15}}
\put(88.00,50.00){\circle{2.00}}
\put(90.00,52.00){\makebox(0,0)[cc]{25}}
\emline{16.00}{89.00}{7}{139.00}{23.00}{8}
\put(16.00,50.00){\circle{2.00}}
\put(17.00,52.00){\makebox(0,0)[cc]{25}}
\put(40.00,38.00){\circle{2.00}}
\put(42.00,40.00){\makebox(0,0)[cc]{3}}
\put(89.00,16.00){\makebox(0,0)[cc]{10}}
\emline{16.00}{50.00}{9}{112.00}{2.00}{10}
\emline{68.00}{132.00}{11}{31.00}{9.00}{12}
\put(88.00,89.00){\circle{2.00}}
\put(90.00,90.00){\makebox(0,0)[cc]{6}}
\put(112.00,37.00){\circle{2.00}}
\put(115.00,38.00){\makebox(0,0)[cc]{3}}
\emline{68.00}{131.00}{13}{127.00}{4.00}{14}
\put(48.00,109.00){\circle{2.00}}
\put(50.00,111.00){\makebox(0,0)[cc]{5}}
\emline{22.00}{149.00}{15}{112.00}{14.00}{16}
\put(18.00,132.00){\circle{2.00}}
\put(21.00,135.00){\makebox(0,0)[cc]{50}}
\put(86.00,12.00){\circle{2.00}}
\emline{86.00}{12.00}{17}{13.00}{134.00}{18}
\put(55.00,63.00){\circle{2.00}}
\put(55.00,58.00){\makebox(0,0)[cc]{36}}
\put(84.00,96.00){\circle{2.00}}
\put(87.00,98.00){\makebox(0,0)[cc]{34}}
\put(31.00,136.00){\circle{2.00}}
\put(34.00,136.00){\makebox(0,0)[cc]{35}}
\put(53.00,85.00){\circle{2.00}}
\put(52.00,88.00){\makebox(0,0)[cc]{40}}
\put(16.00,35.00){\circle{2.00}}
\put(18.00,32.00){\makebox(0,0)[cc]{7}}
\put(16.00,12.00){\circle{2.00}}
\put(18.00,14.00){\makebox(0,0)[cc]{10}}
\put(47.00,111.00){\circle{2.00}}
\put(47.00,114.00){\makebox(0,0)[cc]{11}}
\put(116.00,117.00){\circle{2.00}}
\put(113.00,118.00){\makebox(0,0)[cc]{13}}
\emline{17.00}{50.00}{19}{127.00}{125.00}{20}
\emline{16.00}{89.00}{21}{96.00}{153.00}{22}
\put(49.00,117.00){\circle{2.83}}
\put(50.00,119.00){\makebox(0,0)[cc]{13}}
\emline{16.00}{12.00}{23}{159.00}{88.00}{24}
\emline{16.00}{12.00}{25}{17.00}{149.00}{26}
\put(63.00,39.00){\circle{2.00}}
\put(61.00,41.00){\makebox(0,0)[cc]{30}}
\put(58.00,37.00){\circle{2.00}}
\put(56.00,38.00){\makebox(0,0)[cc]{45}}
\emline{16.00}{89.00}{27}{57.00}{2.00}{28}
\put(49.00,18.00){\circle{2.00}}
\put(53.00,17.00){\makebox(0,0)[cc]{31}}
\put(52.00,15.00){\circle{2.00}}
\put(50.00,9.00){\makebox(0,0)[cc]{38}}
\put(120.00,141.00){\makebox(0,0)[cc]{Fig 16 ($\lambda$, $\beta$) "multiplied by 5".}}
\put(18.00,142.00){\circle{2.00}}
\put(14.00,144.00){\makebox(0,0)[cc]{39}}
\put(16.00,72.00){\circle{2.00}}
\put(12.00,73.00){\makebox(0,0)[cc]{27}}
\put(57.00,135.00){\circle{2.00}}
\put(59.00,137.00){\makebox(0,0)[cc]{4}}
\put(60.00,102.00){\circle{2.00}}
\put(62.00,104.00){\makebox(0,0)[cc]{12}}
\end{picture}

\newpage
\unitlength=1.00mm
\special{em:linewidth 0.4pt}
\linethickness{0.4pt}
\begin{picture}(159.00,158.00)
\emline{11.00}{70.00}{1}{140.00}{70.00}{2}
\emline{140.00}{70.00}{3}{140.00}{141.00}{4}
\put(4.00,13.00){\circle{2.00}}
\put(76.00,13.00){\circle{2.00}}
\put(148.00,13.00){\circle{2.00}}
\put(50.00,44.00){\circle{2.00}}
\put(122.00,44.00){\circle{2.00}}
\put(16.00,115.00){\circle{2.00}}
\put(114.00,9.00){\circle{2.00}}
\emline{114.00}{9.00}{5}{16.00}{116.00}{6}
\emline{148.00}{13.00}{7}{48.00}{133.00}{8}
\put(63.00,115.00){\circle{2.00}}
\put(67.00,113.00){\makebox(0,0)[cc]{20}}
\put(125.00,46.00){\makebox(0,0)[cc]{15}}
\put(141.00,13.00){\makebox(0,0)[cc]{10}}
\put(19.00,115.00){\makebox(0,0)[cc]{6}}
\put(42.00,92.00){\makebox(0,0)[cc]{3}}
\put(113.00,13.00){\makebox(0,0)[cc]{5}}
\put(79.00,15.00){\makebox(0,0)[cc]{10}}
\put(52.00,46.00){\makebox(0,0)[cc]{15}}
\put(8.00,15.00){\makebox(0,0)[cc]{10}}
\emline{76.00}{13.00}{9}{2.00}{101.00}{10}
\put(66.00,24.00){\circle{2.00}}
\put(68.00,25.00){\makebox(0,0)[cc]{30}}
\put(138.00,23.00){\circle{2.00}}
\put(140.00,25.00){\makebox(0,0)[cc]{30}}
\put(23.00,75.00){\circle{2.00}}
\put(27.00,75.00){\makebox(0,0)[cc]{8}}
\put(96.00,75.00){\circle{2.00}}
\put(101.00,75.00){\makebox(0,0)[cc]{8}}
\put(86.00,87.00){\circle{2.00}}
\put(88.00,89.00){\makebox(0,0)[cc]{14}}
\put(111.00,15.00){\circle{2.00}}
\put(112.00,17.00){\makebox(0,0)[cc]{11}}
\put(110.00,12.00){\circle{2.00}}
\put(105.00,9.00){\makebox(0,0)[cc]{19}}
\put(109.00,15.00){\circle{2.00}}
\put(105.00,14.00){\makebox(0,0)[cc]{26}}
\put(97.00,28.00){\circle{2.00}}
\put(100.00,29.00){\makebox(0,0)[cc]{35}}
\put(134.00,115.00){\circle{2.00}}
\put(132.00,117.00){\makebox(0,0)[cc]{20}}
\put(88.00,76.00){\circle{2.00}}
\emline{50.00}{44.00}{11}{143.00}{123.00}{12}
\emline{143.00}{123.00}{13}{16.00}{15.00}{14}
\put(84.00,77.00){\makebox(0,0)[cc]{25}}
\put(128.00,108.00){\circle{2.00}}
\put(130.00,105.00){\makebox(0,0)[cc]{4}}
\put(53.00,27.00){\circle{2.00}}
\emline{76.00}{13.00}{15}{153.00}{65.00}{16}
\put(52.00,21.00){\makebox(0,0)[cc]{40}}
\put(133.00,14.00){\circle{2.00}}
\put(132.00,10.00){\makebox(0,0)[cc]{45}}
\put(90.00,77.00){\circle{2.00}}
\put(90.00,74.00){\makebox(0,0)[cc]{9}}
\put(112.00,59.00){\circle{2.00}}
\put(117.00,58.00){\makebox(0,0)[cc]{16}}
\put(7.00,93.00){\circle{2.00}}
\put(12.00,93.00){\makebox(0,0)[cc]{49}}
\put(19.00,113.00){\circle{2.00}}
\put(22.00,111.00){\makebox(0,0)[cc]{41}}
\put(122.00,2.00){\circle{2.00}}
\put(123.00,4.00){\makebox(0,0)[cc]{43}}
\emline{114.00}{9.00}{17}{123.00}{0.00}{18}
\put(102.00,81.00){\circle{2.00}}
\put(106.00,79.00){\makebox(0,0)[cc]{21}}
\put(80.00,141.00){\makebox(0,0)[cc]{Fig 17 ($\alpha$, $\delta$) "multiplied by 5".}}
\
\put(19.00,18.00){\circle{2.00}}
\put(21.00,15.00){\makebox(0,0)[cc]{2}}
\emline{4.00}{13.00}{19}{159.00}{117.00}{20}
\put(114.00,89.00){\circle{2.00}}
\put(117.00,91.00){\makebox(0,0)[cc]{3}}
\emline{4.00}{13.00}{21}{156.00}{53.00}{22}
\put(23.00,21.00){\circle{2.00}}
\put(27.00,22.00){\makebox(0,0)[cc]{39}}
\put(4.00,156.00){\circle{2.00}}
\put(8.00,152.00){\makebox(0,0)[cc]{50}}
\emline{4.00}{13.00}{23}{4.00}{157.00}{24}
\emline{4.00}{13.00}{25}{148.00}{13.00}{26}
\put(88.00,115.00){\circle{2.00}}
\put(86.00,117.00){\makebox(0,0)[cc]{6}}
\put(149.00,157.00){\circle{2.00}}
\put(149.00,152.00){\makebox(0,0)[cc]{50}}
\emline{23.00}{75.00}{27}{149.00}{157.00}{28}
\put(16.00,60.00){\circle{2.00}}
\put(20.00,57.00){\makebox(0,0)[cc]{7}}
\emline{16.00}{60.00}{29}{117.00}{11.00}{30}
\put(92.00,61.00){\makebox(0,0)[cc]{7}}
\put(88.00,61.00){\circle{2.00}}
\put(42.00,89.00){\circle{2.00}}
\emline{122.00}{44.00}{31}{0.00}{112.00}{32}
\put(59.00,79.00){\circle{2.00}}
\put(63.00,80.00){\makebox(0,0)[cc]{12}}
\put(86.00,67.00){\circle{2.00}}
\put(90.00,66.00){\makebox(0,0)[cc]{48}}
\emline{133.00}{14.00}{33}{4.00}{156.00}{34}
\emline{63.00}{115.00}{35}{156.00}{67.00}{36}
\put(131.00,79.00){\circle{2.00}}
\end{picture}
\newpage
\unitlength=1.00mm
\special{em:linewidth 0.4pt}
\linethickness{0.4pt}
\begin{picture}(144.05,142.00)
\emline{140.00}{70.00}{1}{140.00}{140.00}{2}
\put(63.00,75.00){\circle{0.30}}
\put(66.00,76.00){\makebox(0,0)[cc]{6}}
\put(82.00,45.00){\circle{0.30}}
\put(84.00,41.00){\makebox(0,0)[cc]{2}}
\put(72.00,45.00){\circle{0.30}}
\put(71.00,39.00){\makebox(0,0)[cc]{7}}
\put(100.00,85.00){\circle{0.30}}
\put(101.00,87.00){\makebox(0,0)[cc]{15}}
\emline{140.00}{70.00}{3}{10.00}{70.00}{4}
\put(93.00,71.00){\circle{0.30}}
\put(93.00,71.00){\circle{0.30}}
\put(93.00,73.00){\makebox(0,0)[cc]{5}}
\put(88.00,11.00){\circle{0.30}}
\put(89.00,14.00){\makebox(0,0)[cc]{10}}
\put(104.00,139.00){\circle{0.30}}
\put(107.00,139.00){\makebox(0,0)[cc]{3}}
\put(47.00,72.00){\circle{0.30}}
\put(48.00,74.00){\makebox(0,0)[cc]{20}}
\put(70.00,54.00){\circle{0.30}}
\put(71.00,56.00){\makebox(0,0)[cc]{25}}
\put(29.00,11.00){\circle{0.30}}
\put(28.00,14.00){\makebox(0,0)[cc]{10}}
\emline{29.00}{11.00}{5}{129.00}{115.00}{6}
\put(32.00,139.00){\circle{0.30}}
\put(34.00,142.00){\makebox(0,0)[cc]{3}}
\put(144.00,45.00){\circle{0.30}}
\put(144.00,49.00){\makebox(0,0)[cc]{7}}
\emline{100.00}{85.00}{7}{14.00}{63.00}{8}
\emline{32.00}{139.00}{9}{144.00}{46.00}{10}
\put(134.00,75.00){\circle{0.30}}
\put(135.00,78.00){\makebox(0,0)[cc]{6}}
\emline{135.00}{75.00}{11}{19.00}{107.00}{12}
\put(54.00,97.00){\circle{0.30}}
\put(54.00,99.00){\makebox(0,0)[cc]{50}}
\emline{88.00}{11.00}{13}{54.00}{97.00}{14}
\put(93.00,121.00){\circle{0.30}}
\put(95.00,123.00){\makebox(0,0)[cc]{16}}
\emline{63.00}{75.00}{15}{104.00}{139.00}{16}
\put(89.00,73.00){\circle{0.30}}
\put(87.00,75.00){\makebox(0,0)[cc]{19}}
\put(99.00,18.00){\circle{0.30}}
\put(102.00,14.00){\makebox(0,0)[cc]{30}}
\emline{144.00}{45.00}{17}{77.00}{4.00}{18}
\put(85.00,59.00){\makebox(0,0)[cc]{35}}
\put(83.00,62.00){\circle{0.30}}
\emline{82.00}{45.00}{19}{104.00}{140.00}{20}
\emline{29.00}{11.00}{21}{129.00}{104.00}{22}
\put(99.00,64.00){\circle{0.30}}
\put(101.00,60.00){\makebox(0,0)[cc]{40}}
\emline{104.00}{140.00}{23}{95.00}{15.00}{24}
\emline{88.00}{11.00}{25}{104.00}{139.00}{26}
\emline{72.00}{45.00}{27}{134.00}{75.00}{28}
\put(93.00,55.00){\circle{0.30}}
\put(95.00,52.00){\makebox(0,0)[cc]{43}}
\put(78.00,101.00){\circle{0.30}}
\put(78.00,105.00){\makebox(0,0)[cc]{47}}
\put(59.00,118.00){\circle{0.30}}
\put(62.00,120.00){\makebox(0,0)[cc]{33}}
\emline{32.00}{139.00}{29}{131.00}{29.00}{30}
\put(104.00,60.00){\circle{0.30}}
\put(107.00,59.00){\makebox(0,0)[cc]{36}}
\put(128.00,31.00){\circle{0.30}}
\put(127.00,26.00){\makebox(0,0)[cc]{49}}
\put(69.00,93.00){\circle{0.30}}
\put(68.00,95.00){\makebox(0,0)[cc]{17}}
\put(44.00,71.00){\circle{0.30}}
\put(93.00,93.00){\circle{0.30}}
\put(94.00,87.00){\circle{0.30}}
\put(89.00,88.00){\circle{0.30}}
\emline{82.00}{45.00}{31}{40.00}{109.00}{32}
\emline{40.00}{109.00}{33}{101.00}{16.00}{34}
\put(46.00,99.00){\circle{0.30}}
\put(74.00,57.00){\circle{0.30}}
\put(74.00,60.00){\makebox(0,0)[cc]{32}}
\put(123.00,37.00){\circle{0.30}}
\put(104.00,11.00){\circle{0.30}}
\put(108.00,11.00){\makebox(0,0)[cc]{18}}
\put(107.00,77.00){\circle{0.30}}
\put(102.00,61.00){\circle{0.30}}
\put(103.00,63.00){\makebox(0,0)[cc]{24}}
\put(104.00,84.00){\circle{0.30}}
\put(102.00,81.00){\circle{0.30}}
\put(67.00,66.00){\circle{0.30}}
\put(61.00,63.00){\makebox(0,0)[cc]{27}}
\put(96.00,73.00){\circle{0.30}}
\emline{84.00}{81.00}{35}{139.00}{94.00}{36}
\put(114.00,88.00){\circle{0.30}}
\put(116.00,86.00){\makebox(0,0)[cc]{4}}
\emline{74.00}{92.00}{37}{42.00}{43.00}{38}
\put(55.00,64.00){\circle{0.30}}
\emline{104.00}{139.00}{39}{61.00}{15.00}{40}
\put(65.00,27.00){\circle{0.30}}
\put(74.00,53.00){\circle{0.30}}
\put(77.00,61.00){\circle{0.30}}
\put(76.00,57.00){\circle{0.30}}
\put(100.00,71.00){\circle{0.30}}
\put(135.00,129.00){\makebox(0,0)[cc]{b}}
\put(16.00,72.00){\makebox(0,0)[cc]{l}}
\put(70.00,64.00){\circle{0.30}}
\put(99.00,71.00){\circle{0.30}}
\put(101.00,72.00){\makebox(0,0)[cc]{11}}
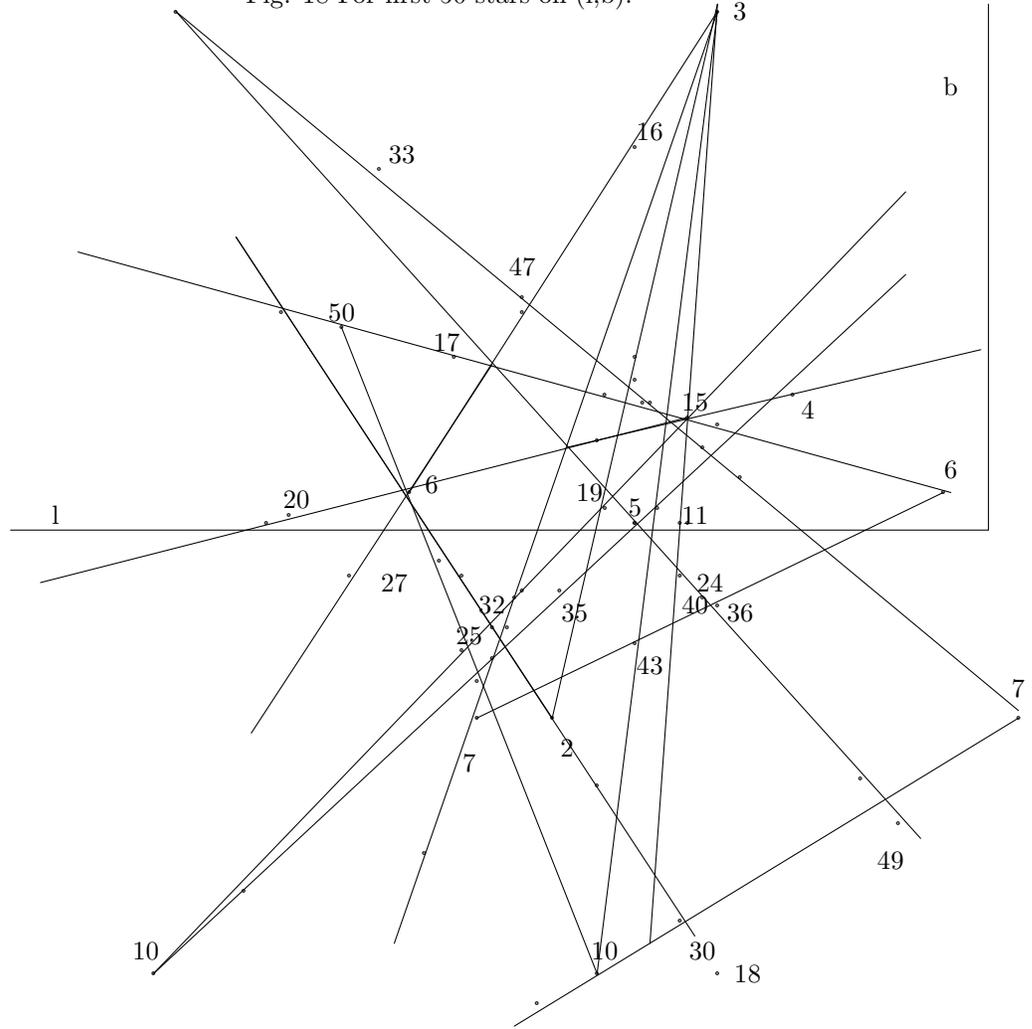
\put(67.00,141.00){\makebox(0,0)[cc]{Fig. 18 For first 50 stars on (l,b).}}
\put(88.00,82.00){\circle{0.30}}
\put(41.00,22.00){\circle{0.30}}
\put(88.00,36.00){\circle{0.30}}
\put(72.00,50.00){\circle{0.30}}
\put(80.00,7.00){\circle{0.30}}
\put(95.00,87.00){\circle{0.30}}
\put(93.00,90.00){\circle{0.30}}
\put(78.00,99.00){\circle{0.30}}
\put(78.00,62.00){\circle{0.30}}
\end{picture}
\newpage
\unitlength=1.00mm
\special{em:linewidth 0.4pt}
\linethickness{0.4pt}
\begin{picture}(153.00,159.18)
\put(50.15,53.42){\circle{0.50}}
\put(49.15,17.33){\circle{1.50}}
\put(72.65,89.45){\circle{1.50}}
\put(85.75,108.73){\circle{0.50}}
\put(73.80,9.37){\circle{1.50}}
\put(45.65,115.95){\circle{1.50}}
\put(45.60,61.75){\circle{1.50}}
\put(52.85,75.35){\circle{0.50}}
\put(47.60,77.40){\circle{0.50}}
\put(34.80,12.52){\circle{1.50}}
\put(72.00,9.87){\circle{1.50}}
\put(89.40,78.73){\circle{0.50}}
\put(67.20,7.18){\circle{0.50}}
\put(43.65,86.42){\circle{0.50}}
\put(79.30,43.68){\circle{1.50}}
\put(70.15,59.10){\circle{0.50}}
\put(53.10,98.15){\circle{0.50}}
\put(98.75,40.12){\circle{0.50}}
\put(68.25,10.85){\circle{0.50}}
\put(92.00,115.10){\circle{0.50}}
\put(60.30,82.22){\circle{0.50}}
\put(50.85,41.10){\circle{0.50}}
\put(52.55,102.00){\circle{0.50}}
\put(82.50,32.93){\circle{0.50}}
\put(46.10,76.30){\circle{0.50}}
\put(67.40,13.17){\circle{0.50}}
\put(46.15,98.57){\circle{0.50}}
\put(57.65,0.48){\circle{0.50}}
\put(46.70,68.77){\circle{0.50}}
\put(96.25,22.80){\circle{0.50}}
\put(68.60,126.23){\circle{0.50}}
\put(46.90,68.03){\circle{0.50}}
\put(63.05,132.02){\circle{0.50}}
\put(40.05,119.68){\circle{0.50}}
\put(54.40,22.80){\circle{0.50}}
\put(85.05,35.58){\circle{0.50}}
\put(51.30,43.68){\circle{0.50}}
\put(71.30,119.57){\circle{0.50}}
\put(55.10,10.65){\circle{0.50}}
\put(82.70,27.03){\circle{0.50}}
\put(47.80,114.95){\circle{0.50}}
\put(49.75,86.45){\circle{0.50}}
\put(80.15,1.07){\circle{0.50}}
\put(56.15,15.48){\circle{0.50}}
\put(91.10,13.10){\circle{0.50}}
\put(49.00,52.07){\circle{0.50}}
\put(58.25,61.57){\circle{0.50}}
\put(36.85,66.80){\circle{0.50}}
\put(36.20,93.23){\circle{0.50}}
\put(35.45,159.03){\circle{0.50}}
\put(87.45,74.00){\circle{0.50}}
\put(30.35,98.95){\circle{0.50}}
\put(86.65,43.68){\circle{0.50}}
\put(65.35,84.70){\circle{0.50}}
\put(70.10,125.05){\circle{0.50}}
\put(74.50,144.25){\circle{0.50}}
\put(39.30,110.85){\circle{0.50}}
\put(36.10,112.20){\circle{0.50}}
\put(30.35,129.00){\circle{0.50}}
\put(31.95,126.00){\circle{0.50}}
\put(33.05,77.00){\circle{0.50}}
\put(68.00,22.00){\circle{0.50}}
\put(65.60,123.00){\circle{0.50}}
\put(72.20,34.00){\circle{0.50}}
\put(74.00,23.00){\circle{0.50}}
\put(73.65,28.00){\circle{0.50}}
\put(70.10,125.00){\circle{0.50}}
\put(74.40,97.00){\circle{0.50}}
\put(76.60,96.00){\circle{0.50}}
\put(78.00,48.00){\circle{0.50}}
\put(80.40,36.00){\circle{0.50}}
\put(83.00,31.00){\circle{0.50}}
\put(83.80,121.00){\circle{0.50}}
\put(91.00,110.00){\circle{0.50}}
\put(99.00,97.00){\circle{0.50}}
\put(93.85,132.47){\circle{0.50}}
\put(81.40,135.00){\circle{0.50}}
\put(82.65,82.57){\circle{0.50}}
\put(32.85,151.57){\circle{0.50}}
\put(34.40,123.75){\circle{0.50}}
\put(34.90,124.70){\circle{0.50}}
\put(35.85,97.63){\circle{0.50}}
\put(37.10,125.87){\circle{0.50}}
\put(38.10,117.93){\circle{0.50}}
\put(38.20,139.42){\circle{0.50}}
\put(39.20,110.75){\circle{0.50}}
\put(40.45,66.53){\circle{0.50}}
\put(42.00,97.98){\circle{0.50}}
\put(43.75,90.58){\circle{0.50}}
\put(44.05,59.23){\circle{0.50}}
\put(45.05,111.22){\circle{0.50}}
\put(45.30,109.52){\circle{0.50}}
\put(45.40,112.10){\circle{0.50}}
\put(46.30,105.33){\circle{0.50}}
\put(46.60,64.58){\circle{0.50}}
\put(46.70,97.23){\circle{0.50}}
\put(47.85,80.22){\circle{0.50}}
\put(49.95,50.38){\circle{0.50}}
\put(50.60,68.70){\circle{0.50}}
\put(56.05,89.20){\circle{0.50}}
\emline{140.00}{70.00}{1}{140.00}{140.00}{2}
\emline{140.00}{70.00}{3}{12.00}{70.00}{4}
\emline{49.00}{17.00}{5}{39.00}{149.00}{6}
\emline{73.00}{89.00}{7}{22.00}{140.00}{8}
\emline{22.00}{140.00}{9}{120.00}{42.00}{10}
\emline{72.00}{10.00}{11}{72.00}{150.00}{12}
\emline{34.00}{13.00}{13}{65.00}{148.00}{14}
\emline{80.00}{43.00}{15}{13.00}{78.00}{16}
\emline{35.00}{12.00}{17}{125.00}{50.00}{18}
\emline{46.00}{62.00}{19}{111.00}{133.00}{20}
\emline{73.00}{90.00}{21}{18.00}{142.00}{22}
\emline{49.00}{18.00}{23}{87.00}{5.00}{24}
\emline{73.00}{9.00}{25}{96.00}{137.00}{26}
\emline{34.00}{12.00}{27}{48.00}{140.00}{28}
\emline{72.00}{10.00}{29}{24.00}{100.00}{30}
\put(118.00,61.00){\circle{2.00}}
\put(152.00,44.00){\circle{2.00}}
\emline{152.00}{44.00}{31}{17.00}{123.00}{32}
\emline{35.00}{13.00}{33}{151.00}{44.00}{34}
\emline{118.00}{61.00}{35}{34.00}{13.00}{36}
\put(77.00,45.00){\circle{0.50}}
\put(77.00,44.00){\circle{0.50}}
\put(78.00,45.00){\circle{0.50}}
\emline{79.00}{43.00}{37}{64.00}{147.00}{38}
\emline{64.00}{147.00}{39}{84.00}{10.00}{40}
\put(78.00,48.00){\circle{0.50}}
\put(78.00,52.00){\circle{0.50}}
\put(77.00,53.00){\circle{0.50}}
\put(78.00,52.00){\circle{0.50}}
\put(74.00,28.00){\circle{0.50}}
\put(75.00,29.00){\circle{0.50}}
\put(77.00,29.00){\circle{0.50}}
\put(73.00,27.00){\circle{0.50}}
\put(72.00,26.00){\circle{0.50}}
\put(76.00,23.00){\circle{0.50}}
\put(77.00,24.00){\circle{0.50}}
\emline{34.00}{13.00}{41}{99.00}{144.00}{42}
\put(46.00,37.00){\circle{0.50}}
\put(48.00,35.00){\circle{0.50}}
\put(48.00,32.00){\circle{0.50}}
\put(45.00,34.00){\circle{0.50}}
\put(37.00,76.00){\circle{0.50}}
\put(38.00,74.00){\circle{0.50}}
\put(40.00,71.00){\circle{0.50}}
\put(35.00,78.00){\circle{0.50}}
\emline{46.00}{116.00}{43}{46.00}{0.00}{44}
\put(46.00,72.00){\circle{0.50}}
\put(45.00,76.00){\circle{0.50}}
\put(46.00,96.00){\circle{0.50}}
\put(46.00,92.00){\circle*{0.50}}
\put(78.00,101.00){\circle*{2.00}}
\put(87.00,118.00){\circle*{2.00}}
\put(97.00,138.00){\circle*{2.00}}
\put(100.00,144.00){\circle*{2.00}}
\put(46.00,112.00){\circle*{2.00}}
\put(45.00,114.00){\circle*{2.00}}
\put(46.00,106.00){\circle*{2.00}}
\put(47.00,100.00){\circle*{2.00}}
\put(68.00,86.00){\circle*{2.00}}
\put(68.00,86.00){\circle*{2.00}}
\put(82.00,101.00){\circle*{2.00}}
\put(90.00,109.00){\circle*{2.00}}
\put(89.00,109.00){\circle*{2.00}}
\put(89.00,109.00){\circle*{2.00}}
\put(89.00,97.00){\circle*{2.00}}
\put(89.00,96.00){\circle*{2.00}}
\put(87.00,90.00){\circle*{2.00}}
\put(89.00,93.00){\circle*{2.00}}
\put(88.00,88.00){\circle*{2.00}}
\put(88.00,88.00){\circle*{2.00}}
\emline{72.00}{10.00}{45}{102.00}{143.00}{46}
\put(93.00,101.00){\circle*{2.00}}
\put(96.00,112.00){\circle*{2.00}}
\put(44.00,119.00){\circle*{2.00}}
\put(39.00,123.00){\circle*{2.00}}
\put(33.00,130.00){\circle*{2.00}}
\put(25.00,137.00){\circle*{2.00}}
\put(19.00,142.00){\circle*{2.00}}
\put(103.00,125.00){\circle*{2.00}}
\put(108.00,130.00){\circle*{0.00}}
\put(108.00,131.00){\circle*{2.00}}
\put(72.00,99.00){\circle*{2.00}}
\put(70.00,107.00){\circle*{2.00}}
\put(109.00,56.00){\circle*{2.00}}
\put(96.00,76.00){\circle*{2.00}}
\put(96.00,76.00){\circle*{2.00}}
\put(109.00,69.00){\circle*{2.00}}
\put(118.00,115.00){\circle{0.50}}
\emline{72.00}{90.00}{47}{146.00}{131.00}{48}
\put(112.00,112.00){\circle*{2.00}}
\put(116.00,113.00){\circle*{2.00}}
\put(93.00,104.00){\circle*{2.00}}
\put(42.00,111.00){\circle*{2.00}}
\put(41.00,130.00){\circle*{2.00}}
\put(39.00,132.00){\circle*{2.00}}
\put(43.00,91.00){\circle*{2.00}}
\put(44.00,81.00){\circle*{2.00}}
\put(47.00,69.00){\circle*{2.00}}
\put(92.00,127.00){\circle*{2.00}}
\put(96.00,134.00){\circle*{2.00}}
\put(35.00,23.00){\circle{0.50}}
\put(36.00,26.00){\circle{0.50}}
\put(77.00,34.00){\circle{0.50}}
\put(72.00,19.00){\circle{0.50}}
\put(72.00,24.00){\circle{0.50}}
\put(72.00,22.00){\circle{0.50}}
\put(65.00,13.00){\circle{0.50}}
\put(73.00,76.00){\circle{0.50}}
\put(62.00,29.00){\circle{0.50}}
\put(130.00,21.00){\makebox(0,0)[cc]{Fig 19 For first light 500 stars on $\alpha, \beta$. }}
\put(106.00,10.00){\circle*{2.00}}
\put(125.00,10.00){\makebox(0,0)[cc]{EA, EB stars}}
\put(43.00,15.00){\circle{0.50}}
\put(56.00,55.00){\circle{0.50}}
\put(57.00,58.00){\circle{0.50}}
\put(63.00,80.00){\circle{0.50}}
\put(79.00,38.00){\circle{0.50}}
\put(47.00,88.00){\circle{0.50}}
\put(116.00,65.00){\circle{0.50}}
\put(43.00,65.00){\circle{0.50}}
\put(42.00,67.00){\circle{0.50}}
\emline{46.00}{116.00}{49}{98.00}{3.00}{50}
\put(76.00,50.00){\circle{0.50}}
\put(77.00,49.00){\circle{0.50}}
\put(75.00,52.00){\circle{0.50}}
\emline{46.00}{116.00}{51}{32.00}{155.00}{52}
\put(49.00,110.00){\circle{0.50}}
\emline{152.00}{44.00}{53}{46.00}{1.00}{54}
\emline{118.00}{61.00}{55}{61.00}{0.00}{56}
\put(88.00,28.00){\circle{0.50}}
\put(90.00,30.00){\circle{0.50}}
\put(85.00,26.00){\circle{0.50}}
\put(99.00,40.00){\circle{0.50}}
\put(101.00,43.00){\circle{0.50}}
\put(98.00,40.00){\circle{0.50}}
\put(97.00,37.00){\circle{0.50}}
\put(78.00,19.00){\circle{0.50}}
\put(78.00,16.00){\circle{0.50}}
\put(94.00,12.00){\circle{0.50}}
\end{picture}

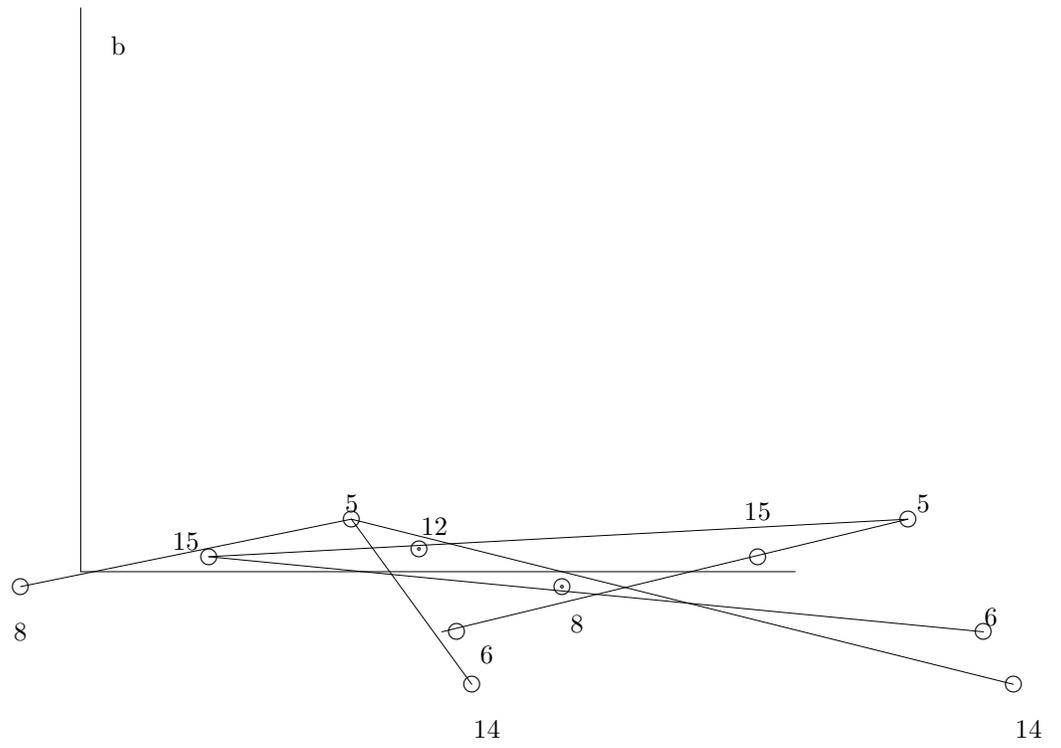
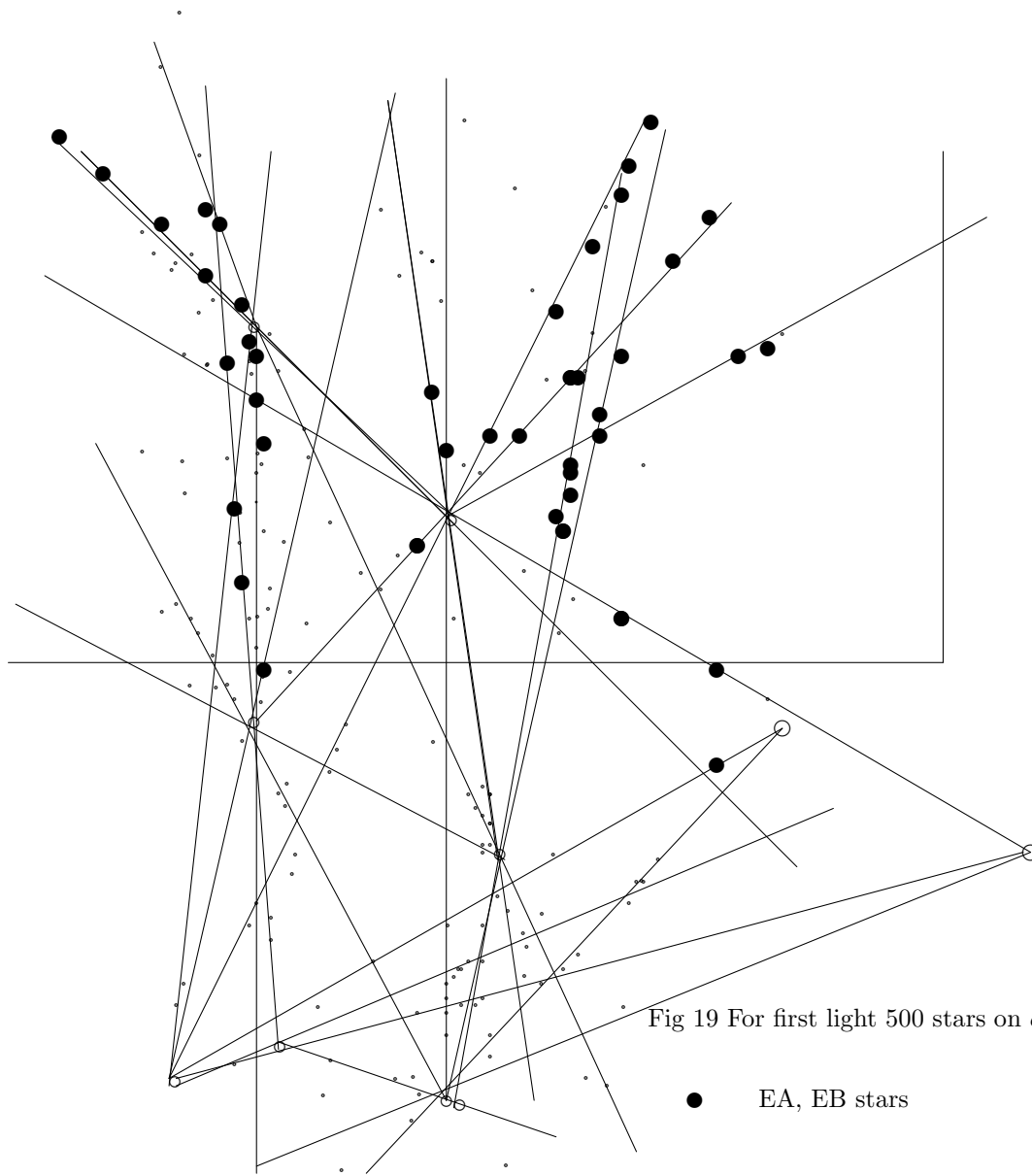
\end{document}